\def\year{2020}\relax
\documentclass[letterpaper]{article} 
\usepackage{aaai20}  
\usepackage{times}  
\usepackage{helvet} 
\usepackage{courier}  
\usepackage[hyphens]{url}  
\usepackage{graphicx} 
\usepackage[table,xcdraw]{xcolor}
\usepackage{graphicx}  
\title{edBB: Biometrics and Behavior for Assessing Remote Education}
\author{Javier Hernandez-Ortega, Roberto Daza, Aythami Morales, Julian Fierrez, Javier Ortega-Garcia\\
BiDA Lab., Universidad Autonoma de Madrid, Spain\\ 
javier.hernandezo@uam.es, roberto.daza@uam.es, aythami.morales@uam.es, julian.fierrez@uam.es, javier.ortega@uam.es 
}
 
\begin{document}

\maketitle

\begin{abstract}

We present a platform for student monitoring in remote education consisting of a collection of sensors and software that capture biometric and behavioral data. We define a collection of tasks to acquire behavioral data that can be useful for facing the existing challenges in automatic student monitoring during remote evaluation. Additionally, we release an initial database including data from $20$ different users completing these tasks with a set of basic sensors: webcam, microphone, mouse, and keyboard; and also from more advanced sensors: NIR camera, smartwatch, additional RGB cameras, and an EEG band. Information from the computer (e.g. system logs, MAC, IP, or web browsing history) is also stored. During each acquisition session each user completed three different types of tasks generating data of different nature: mouse and keystroke dynamics, face data, and audio data among others. The tasks have been designed with two main goals in mind: i) analyse the capacity of such biometric and behavioral data for detecting anomalies during remote evaluation, and ii) study the capability of these data, i.e. EEG, ECG, or NIR video, for estimating other information about the users such as their attention level, the presence of stress, or their pulse rate. 
\end{abstract}

\section{Introduction}

\noindent In recent years virtual education (e-learning) and electronic evaluation is undergoing a process of constant expansion, and it is expected to continue growing in the near future. \cite{santamans2014mercado} estimated that the virtual education industry will grow over a $5$\% between $2018$ and $2023$, reaching a turnover around $240,000$ million dollars compared to the $56,000$ millions obtained in $2013$. This significant evolution is undoubtedly influenced by the need of a continuous learning process during the whole professional life, specially in the case of technological careers in which the state-of-the-art is continuously advancing. In many cases the traditional scheme of face-to-face education is not an option due to time limitations, geographical constraints, etc. 

A survey conducted in $2016$ to $25,000$ people worldwide and published in the World Economic Forum \cite{WEFsurvey2016}, reflected the relevance of virtual education, with over a $78$\% of the participants claiming to have taken at least one online course in their lifetime. The survey also showed that almost half of the participants had plans to continue looking for online courses once they had started their working activity. 

As a consequence of its large expansion, online education has become an integral part of the general strategy of higher education institutions. In \cite{bowers2015students} it is shown that in $2015$, over a $90$\% of higher education centers offered some kind of online education, including some of the most relevant universities in the world such as Stanford University and Oxford University. 

Virtual education offers great flexibility, e.g. giving students the ability to connect to the teaching platform at any time and any place, compared with traditional education that establishes strict schedules and mandatory physical attendance. Thanks to this flexibility, students with an Internet connection have the possibility of studying independently of their location and schedule. Virtual education also permits a higher number of people to access the same training contents than when using a face-to-face alternative, that usually requires a physical space, and more teaching personnel. However, it does not guarantee a more effective or faster learning process compared to traditional teaching. Additionally, problems may appear derived from the dependence of a functional Internet connection, a device, or of a teaching platform. The learning process can also be harder for the students due to the distant relation between them and the teachers. 

Among all the possible drawbacks of virtual education, one of them is being constantly remarked by those who question it: the actual capability of current technology to demonstrate that an online evaluation is being really carried out by a specific student without incurring in any type of fraud or cheating. Due to this point, many of the educational institutions choose to perform online teaching but face-to-face evaluation, loosing in this way a large part of the benefits that virtual education can generate and increasing the final costs.

It is at this point that biometric technologies appear as an alternative for student authentication in the virtual evaluation process. These technologies permit to identify a person by their physiological and behavioral characteristics, rather than by other methods such as a password or an ID card that could be used by another person for performing student impersonation \cite{hadid15SPMspoofing}.

The contributions of this work are: i) we present a survey of state-of-the-art biometric and behavioral technologies with potential applicability to student monitoring (based on Human-Computer Interaction); ii) we present a student monitoring platform for e-learning consisting of a collection of sensors and software that capture state-of-the-art biometrics and behavioral data; and iii) we have designed a series of tasks associated to $5$ challenges that served us to collect a database that will be available for the research community\footnote{Available at GitHub: https://github.com/BiDAlab/edBBdb}; iv) we have released an initial subset of that database including signals from $16$ sensors captured from $20$ different users.   

The rest of this paper is organized as follows: Section \ref{Related} summarizes works related to trustworthy online evaluation. Section \ref{behavioral} introduces Behavioral Biometrics and their application to virtual evaluation. Section \ref{system} describes the proposed monitoring platform, i.e. its architecture, and the different acquisition tasks. Section \ref{challenges} explains the proposed evaluation challenges and the acquired database. Finally, concluding remarks and future work are drawn in Section \ref{conclusion}.

\section{Related Works}
\label{Related}

A number of solutions have been recently proposed for reliable and trustworthy online evaluations. A first approach consists in using a special software installed on the students' computer that is connected through the Internet to the Learning Management System (LMS) of the educational institution. This type of software is responsible of controlling that the student does not execute unauthorized actions during the evaluation, e.g. opening restricted applications, web connections, closing the evaluation software before completing the exam, using keyboard shortcuts, making screenshots, etc. This software usually offers encryption mechanisms to guarantee the privacy and security of students. Examples of these services are Safe Exam Browser \cite{mondal2017study} (a browser based on a lite version of Firefox), and Secure Exams (a browser that protects the integrity of any web-based exam). However, from a pedagogical point of view, this type of control may have negative effects on students, possibly leading to high stress levels that can affect the test results \cite{bailey2014user,martin2016strokes}. In addition, this type of approach has the drawback of not being able to verify the identity of the student that is taking the exam/evaluation.

Another option is the use of online supervisors. This approach allows students to take exams from their computer while being monitored in real time by a remote supervisor using tools like the webcam, the microphone, desktop monitoring, etc. The biggest drawback of this alternative is its lack of scalability, since supervisors are needed in real time; the lack of privacy (since most of the time supervisors are people far from the academic field, coming from an external company); and its high cost. Some examples of these online services are Kryterion, ProctorU, and Pearson VUE.

The previous alternatives do not guarantee a reliable online evaluation (due to the lack of authentication) and scalability at the same time. At this point, biometric technologies arise as a solution to create a reliable and scalable online evaluation system. An early proof of their potential can be seen in the Coursera platform, which uses biometric traits such as keystroke dynamics \cite{morales2016keystroke} for authenticating students who are enrolled in their official certifications.

One of the most important projects for developing an online evaluation platform using biometric technologies is the EU project called TeSLA, which is currently in its testing phase. Identification and authentication in TeSLA is performed using keystroke dynamics, facial recognition, and voice recognition among other traits. The results obtained by TeSLA in the preliminary tests performed at the Technical University of Sofia are very favorable \cite{baro2018integration,ivanova2019enhancing}, achieving an Equal Error Rate (EER) of $1.1$\% when using face recognition, an EER of $8.85$\% with voice recognition, and an EER ca. $2$\% for keystroke biometrics.


\section{Behavioral Biometrics for Student Monitoring based on HCI}
\label{behavioral}

Behavioral biometrics refers to those biometric traits revealing distinctive user behaviors and mannerisms when performing specific actions \cite{jain201650}. Behavioral biometrics characteristics can be acquired almost transparently to the users, being less invasive than other methodologies, thanks to Human-Computer Interaction (HCI) \cite{salah2013understanding,shrobe2018behavioral}.

The interaction between humans (the users) and computers generates data with patterns affected by: human characteristics (e.g. attitude, emotional state, neuromotor and cognitive abilities); sensor characteristics (e.g. ergonomics, precision), and task characteristics (e.g. easy of use, design, usefulness). Modelling the user behavior using information from heterogeneous data is an ongoing challenge with applications in a variety of fields such as security, e-health, gaming, and education. Behavioral biometrics have been used to model inner human features like cognition or motor cortex skills. The literature of behavioral biometrics is large and includes different traits like keystroking, mouse dynamics \cite{chen2001can}, handwriting patterns \cite{2017_PLOSONE_eBioSign_Tolosana}, touchpad interaction, and stylometry. Here we briefly describe the main data streams that can be considered for monitoring the student behavior based on biometric information:


\begin{itemize}

    \item \textbf{Camera and Microphone}: These are two of the most common sensors in HCI, and also for user authentication. They can take photos, videos, or record voice and sounds. However, their capacity to collect personal information without the user being aware can be perceived as intrusive sometimes.  

    \item \textbf{Keystroke dynamics}: Keystroke data is widely used since it is easy to acquire in a transparent setup. Results are promising in challenging tasks such as user recognition using either free text (i.e. typing any kind of text \cite{tappert2012keystroke}), or fixed text \cite{morales2016keystroke} (i.e. typing a prefixed text like passwords). 
    
    
    \item \textbf{Stylometry}: Stylometry is defined as the study of the linguistic styles of users in order to determine the authorship of texts. In \cite{stewart2011investigation,locklear2014continuous} the authors proposed stylometry-based features to improve keystroke user recognition. The results suggested that while keystroke biometrics operates at an automatic neuromotor control level, stylometry biometrics operates at a higher cognitive level where both words and syntax-level units are involved.

    \item \textbf{Mouse dynamics and gaze}: Mouse dynamics are affected by the specific neuromotor characteristics of each user. In \cite{ahmed2007new} and \cite{gamboa2007webbiometrics} researchers explored features obtained from mouse tasks for user recognition. Their results achieved up to $95$\% of authentication accuracy. Besides, mouse dynamics are usually combined with keystroke information in continuous authentication schemes \cite{mondal2017study,2018_INFFUS_MCSreview1_Fierrez}. In addition, in \cite{chen2001can}, the authors studied the relationship between eye gaze position and mouse cursor position on a computer screen during web browsing, and suggested that there exist regular patterns of eye/mouse movements associated to the characteristics of the motor cortex system of each user.
    
    
    \item \textbf{Wearables}: In recent years, the use of wearable devices has proliferated in people's daily lives \cite{hill2015wearables}. This category includes smartwatches, smartbands, and clothes with sensors, as well as smartphones, since these ones are often carried by users inside their pockets or handbags. Thanks to their popularization, obtaining continuous data about the users' activity and physiology is possible almost at any time. The data available in this case include: GPS location, gyroscope, accelerometer, heart rate data, blood pressure, body temperature, etc. Knowing these metrics, many other parameters can be estimated or even predicted, such as the level of attention, stress, or users' vigilance \cite{hammerla2016deep}.

\end{itemize}

Additionally to the sensors that capture biometric information, in HCI there are other data that can help to know more about the users, their activity, and the computers they are using:

\begin{itemize}

    \item \textbf{Digital Context}: This category includes information about the usage of applications, e.g. the web browser, and also information about the computer itself (MAC adress, IP address, system logs, etc.) 
    
    
    Computer information such as system logging, or the IP and MAC addresses can be useful for detecting impersonation attacks or other hacking techniques. These attacks can be performed both from student's side and also externally, and they can be used for cheating in an online evaluation, and also for accessing to sensitive information.

    \item \textbf{Screen Monitoring}: The information being displayed on the screen can be captured at any moment in a transparent manner. This data can be used to know if an application is opened/closed, or which task is actually being performed at the computer. This knowledge can be useful for monitoring students during a virtual evaluation in which the use of some applications may be forbidden or restricted. The screen data can be also correlated with other information like eye-gaze, head pose, and mouse tracking among others \cite{cheung2015eye}.
    
\end{itemize}

The  behavioral biometrics literature demonstrates that most of the previous sensors and data streams can be used not only for authentication in the virtual evaluation scenario, but also for modelling other human features (e.g. cognitive functions, neuromotor skills, physiological signals, and human behaviors/routines) during the interaction. 


\begin{table*}[t!]
\begin{center}
\resizebox{\linewidth}{!}{
\begin{tabular}{|c|c|c|l|}

\cline{1-4}
\cellcolor[HTML]{C0C0C0}{\color[HTML]{000000} \textbf{Information Type}} & \cellcolor[HTML]{C0C0C0}{\color[HTML]{000000} \textbf{Sensors}}                                                         & \cellcolor[HTML]{C0C0C0}{\color[HTML]{000000} \textbf{Sampling Rate}} & \multicolumn{1}{c|}{\cellcolor[HTML]{C0C0C0}{\color[HTML]{000000} \textbf{Features}}}\\ 

\cline{1-4}
\textbf{Video} & \begin{tabular}[c]{@{}c@{}} 4 RGB cameras\\ 2 Infrared cameras\\ 1 Depth camera\end{tabular} & 20 Hz - 30 Hz & \begin{tabular}[c]{@{}l@{}}MP4 files with codec H264\end{tabular}\\ 

\cline{1-4}
\textbf{\begin{tabular}[c]{@{}c@{}}Desktop Video\end{tabular}}  & Screen  & \begin{tabular}[c]{@{}c@{}}1 Hz\end{tabular} & \begin{tabular}[c]{@{}l@{}}MP4 file with codec H264\end{tabular} \\ 

\cline{1-4}
\textbf{Audio} & Microphone  & \begin{tabular}[c]{@{}c@{}}8000 Hz\end{tabular}  & \begin{tabular}[c]{@{}l@{}}Uncompressed WAV files\end{tabular} \\ 

\cline{1-4}
\textbf{Keystroke}    & Keyboard  & 12 Hz & \begin{tabular}[c]{@{}l@{}}Keyboard events:\\ Keypress and key release events\end{tabular} \\ 

\hline
\textbf{\begin{tabular}[c]{@{}c@{}}Mouse Dynamics\end{tabular}} & Mouse  & 895 Hz  & \begin{tabular}[c]{@{}l@{}}Mouse events:\\   Move, press/release, drag and drop and mouse wheel spin\end{tabular} \\ 

\hline
\textbf{EEG} & Band  & \begin{tabular}[c]{@{}c@{}}1 Hz\end{tabular}   & \begin{tabular}[c]{@{}l@{}}  Power Spectrum Density of five frequency bands. \\   Level of attention (from 0 to 100) and eye blink strength\end{tabular} \\ 

\hline
\textbf{\begin{tabular}[c]{@{}c@{}}Pulse and Inertial\end{tabular}} & \begin{tabular}[c]{@{}c@{}}SmartWatch: \\PPG, Gyroscope\\ Accelerometer, Magnetometer\end{tabular} & \begin{tabular}[c]{@{}c@{}}200 Hz\end{tabular} & \begin{tabular}[c]{@{}l@{}}Timestamps and data from the pulse and \\the inertial sensors (accelerometer, gyroscope, magnetometer)\end{tabular} \\ 

\hline
\textbf{Context Data}   & \begin{tabular}[c]{@{}c@{}}Student, Computer, Server\end{tabular}  & NA  & \begin{tabular}[c]{@{}l@{}}Computer name, private IP, public IP, MAC, OS, architecture \\ keyboard language, screen resolution, free memory, \\ main memory, start time and finish time of the test \\ time in each homework, test answers\end{tabular} \\ 

\hline
\end{tabular}
}
\caption{\textbf{Sensors and data} captured by the evaluation monitoring system. In this paper, in order to acquire a dataset of biometric and non biometric data, each student has completed several challenges/tasks.}
\label{tab:sensors}
\end{center}
\end{table*}

\section{edBBplat: A Platform of Biometrics and Behavior for Remote Education}
\label{system}

    

\subsection{Sensors}
\label{sensors}
Table \ref{tab:sensors} shows the sensors and the types of data captured by the platform. The data was acquired after designing a set of activities for the users to complete.

The acquisition setup consisted of the next components (see Figure \ref{data}):
\begin{itemize}

    \item $3$ individual \textbf{RGB cameras} (frontal, side, and cenital), and $1$ \textbf{Intel Real-Sense} (model D435i), which is composed by $1$ RGB and $2$ Near Infrared sensors, and which also computes also depth images combining its $3$ image channels.

    \item A \textbf{Huawei Watch 2} that captures pulse information in real time and has also accelerometer, magnetometer, and gyroscope; useful to measure the arm movements. 

    \item An \textbf{EEG headset} by NeuroSky that captures $3$ channels of electroencephalogram information. These data can be employed to know the focus level, stress, vigilance, etc. of the students.

    \item A \textbf{Personal Computer} with Microsoft Windows 10 OS, a microphone to acquire audio, a regular keyboard, a mouse, and a screen. The computer is employed both to complete the tasks and also to acquire the screen data, the mouse and keyboard dynamics, audio information during the evaluation, and several types of metadata (e.g. logging, app and web history, IP and MAC addresses, etc.)

\end{itemize}

A summary of all the sensors and the types of information that have been acquired can be seen in Table \ref{tab:sensors}. We have divided the sensors in three categories according to their availability in common remote education scenarios:

\begin{itemize}
    \item \textbf{Basic sensors:} Frontal camera, keystroke, mouse, screen, and microphone. These are sensors typically available in a traditional desktop computer setup.
    
    \item \textbf{Advanced sensors:} IR cameras, smartwatch, EEG band. They provide biometric signals related with the emotional state of the student. These sensors are not normally available in a traditional setup.

    \item \textbf{Extra sensors:} cenital and side cameras. They provide a general outlook of the scene.
\end{itemize}

\subsection{Tasks}
\label{tasks}
The activities designed to conform the database consist of $8$ different tasks that can be categorized in the following three groups:

\begin{itemize}
    \item \textbf{Enrollment form}: Its target consists in obtaining personal data of the users such as their name and surname, ID number, nationality, e-mail address, etc. This form is designed to acquire different events such as the mouse dynamics, clicks, mouse wheel, keyboard use, etc. 
    
    \item \textbf{Writing questions}: These comprehend questions that require a complex interaction from the user. They are oriented to measure the students' cognitive abilities under different situations such as: solving logical problems, describing images, crosswords, finding differences, etc. Additionally, some activities have been designed to induce different states of emotions to the participants, e.g. stress or nervousness. These altered states are highly relevant when working with physiological and biological signals.
    
    \item \textbf{Multiple choice questions}: These are questions aimed to detect the students' attention and focus levels. Since multiple choice exams are largely used in online assessment platforms to evaluate their students, including these in our evaluation was essential. 
    
\end{itemize}

The questions are selected from popular riddles and they present different levels of difficulty. The interface is designed to ensure data from different nature: free text typing (writing questions), fixed text typing (enrollment form), mouse movement (multiple choice questions), visual attention (describing images and finding differences), etc.


\section{Database and Challenges}
\label{challenges}

The initial subset of the full database that is released with the present paper is composed by 20 users captured under controlled laboratory conditions during one session\footnote{Available at GitHub: https://github.com/BiDAlab/edBBdb}. The enrollment form includes demographic information from the user (age, gender, right-handed or left-handed). Additionally, we provide the performance (accuracy and time) achieved by each user in each specific task. Together with the raw data obtained from the sensors, the database includes information processed to better understand and model the student behavior. This information is obtained using state-of-the-art algorithms:

\begin{itemize}
    
    \item \textbf{Head Pose:} head pose (pitch, roll, and yaw) is estimated from the frontal webcam using the algorithm proposed in \cite{ruiz2018cvpr}. 
    
    \item \textbf{Mental State:} attention and meditation is estimated from the EEG signals according to the method developed by NeuroSky. The attention indicates the intensity of mental “focus”.  The value ranges from 0 to 100. The attention level increases when a student focuses on a single thought or an external object, and decreases when distracted. The meditation indicates the level of mental relaxation. The value ranges from 0 to 100, and increases when users relax the mind and decreases when they are uneasy or stressed. 
    
    \item \textbf{Face Biometrics:} size of the face (related to the distance to the front webcam) and authentication score are provided using the face detection algorithm proposed in \cite{Zhang2016signal} and the face authentication model \cite{cao2018vggface2}.
    
\end{itemize}

\begin{figure*}[t]
\centering
\includegraphics[width=1\textwidth]{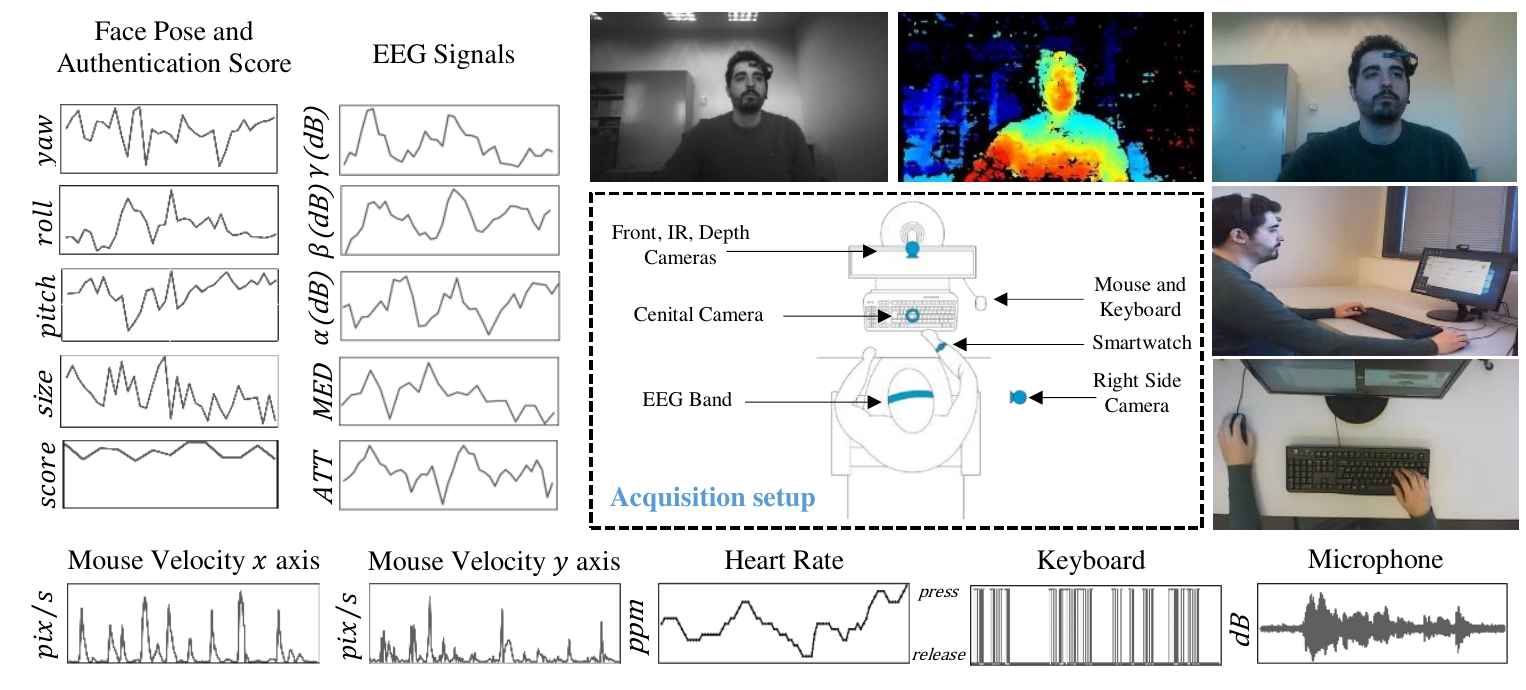} 
\caption{Example of the information captured with the edBB platform. Note that not all signals are included in the figure. Sensors included in the figure: front webcam, IR camera, depth camera, side and zenital camera, head pose features estimated from the frontal webcam, EEG signal obtained from the band, heart rate from the smartwatch, mouse velocities, keyboard events, and microphone signal.}
\label{data}
\end{figure*}

Figure \ref{data} shows an example of the information captured during the execution of the tasks.

We have designed an acquisition protocol incorporating all sensors presented in Section \ref{sensors} and tasks presented in Section \ref{tasks}. Some of the sensors are used to capture the groundtruth for the different challenges proposed. We propose $5$ challenges related to the monitorization of different behaviors relevant for e-learning platforms. For each challenge, we propose target and input data. The goal is to train new artificial intelligence models capable of predicting the target from the input data. The $5$ challenges proposed are (also available in the GitHub mentioned before):  

\begin{itemize}
    
    \item \textbf{Challenge $\textbf{1}$ - Attention Estimation:} an estimation of the attention level of the students during the execution of e-learning tasks is a very valuable resource. We propose to estimate the band signals (level of attention) from patterns captured from the basic sensors. The head pose and gaze estimation from the webcam, together with the mouse and keystroke dynamics can be used to predict attention of the students. \textbf{\textit{Target:}} attention level obtained from the band signals. \textbf{\textit{Input:}} front webcam video, mouse, and keystroke sequences.   

    \item \textbf{Challenge $\textbf{2}$ - Anomalous Behavior Detection:} the detection of non-allowed behaviors during the execution of evaluation tasks is an important challenge necessary to improve the trustworthiness in e-learning platforms. Ten users were instructed to perform non-allowed activities during the execution of the tasks. These activities comprise the use of material/resources with the correct responses to the questions. We propose the use of a smartphone as a non-allowed resource. These users try to hide the smartphone in their pockets. These events are labelled with a timestamp that identify the exact period when cheating really occurred.. We propose to use the basic sensors to detect these events. \textbf{\textit{Target:}} detection of non-allowed events. \textbf{\textit{Input:}} front webcam video, microphone, mouse, and keystroke dynamics. 
    
    \item \textbf{Challenge $\textbf{3}$ - Performance Prediction:} each task is evaluated and the performance is measured in terms of accuracy (percentage of correct responses) and time spent to complete the task. We propose to estimate the performance of the student using both basic and advance sensors. \textbf{\textit{Target:}} accuracy. \textbf{\textit{Basic Input:}} front webcam video, mouse, and keystroke. \textbf{\textit{Advanced Input:}} basic sensors plus pulse and EEC band signals.  

    \item \textbf{Challenge $\textbf{4}$ - User Authentication:} student authentication is a critical step in a e-learning platforms.  All users complete the same tasks, including the enrollment form that contains personal data. Data is anonymized but an ID number is provided to identify data from each user. The dataset is rich in biometric patterns useful for authentication (face, keystroke, mouse). \textbf{\textit{Target:}} identity of the student. \textbf{\textit{Basic Input:}} front webcam video, mouse, and keystroke dynamics. \textbf{\textit{Advanced Input:}} IR cameras, smartwatch sensors, EEG band.
    
    \item \textbf{Challenge $\textbf{5}$ - Pulse Estimation:} the pulse is highly related to the emotional state and stress level of the students. In this challenge, we propose to estimate the pulse from the smartwatch using the front camera. Alternatively, the IR cameras can be used to analyse the potential of these sensors. \textbf{\textit{Target:}} pulse of the student. \textbf{\textit{Basic Input:}} front webcam video. \textbf{\textit{Advanced Input:}} IR cameras.  

\end{itemize}


\section{Conclusion and Future Work}
\label{conclusion}
In this paper, we have: i) discussed the application of behavioral biometrics for remote education, ii) presented a platform of biometrics and behavior for this application, iii) designed a series of tasks and challenges for acquiring biometric and behavioral data including HCI information during virtual evaluations, and iv) presented a subset of the database currently being captured with data both from basic sensors (those typically present in remote education), and also from advanced sensors (e.g. NIR camera, smartwatch, EEG band, etc.) 

Biometric and Behavior information can be used to look for anomalies during the evaluation process, e.g. attempts to cheat, and also to extract other information from the users such as their stress level, their attention level, or even their pulse rate.

The configuration of the acquisition setup consisted of the sensors that are usually found during an online evaluation: a RGB webcam, a microphone, a mouse, a keyboard, and the computer. We also added some advanced sensors not present so often: a smartwatch, an EEG band, a NIR camera, and also additional RGB cameras. 

The released subset of data contains tasks performed by $20$ users during one single session. During each acquisition session, the users had to complete different tasks: an enrollment form, multiple choice questions, and writing questions. Each one of these tasks is designed to capture different information from users such as mouse dynamics, keystroke dynamics, face data, audio, or EEG data.

Student monitoring during virtual evaluation presents several challenges, including: the detection of anomalies that may happen during evaluation, attention estimation, anomalous behavior detection, performance prediction, user authentication, and pulse estimation. We have designed our data acquisition with all these challenges in mind, for example by instructing the users to perform non-allowed activities during the tasks (useful for anomalous behavior detection).

For future work, we expect to extend the database with more users, tasks, and acquisition sessions. Adding different tasks and challenges will help us to detect other types of anomalies, and also to get more information about those currently defined.

Additionally, we are developing a framework for automatically checking the presence of anomalies in the data from a real evaluation, and generating a final report. This report will highlight if anything is happening during the evaluation and when it is occurring. This report will help the review process of virtual evaluations, increasing the reliability of the process and reducing the need for reviewing staff, saving time and costs.

Finally, we also plan to employ the data from both the basic and advanced sensors for studying how other factors such as the stress level, the focus level, or the pulse rate affect to the students' performance during their evaluation. This information may be also useful for its application in other research topics such as health monitoring \cite{2019_FG_PDhandw_Castrillon}.

\section{ Acknowledgments}
\label{ack}
This work has been supported by projects BIBECA (RTI2018-101248-B-I00 MINECO/FEDER) and Bio-Guard (Ayudas Fundacion BBVA a Equipos de Investigacion Cientifica 2017); and by Universidad Autonoma de Madrid.

\bibliographystyle{aaai}
\bibliography{egbib}

\end{document}